\documentclass[onecolumn,12pt]{article}
\title{Modeling of aging processes in the insertion compounds}

\author{E.V. Vakarin, J.P. Badiali\\
\vspace{0.3cm}
LECIME (UMR 7575), CNRS-ENSCP,\\ 11 rue P. et M. Curie, 75231 Cedex 05, Paris, France}

\date{}

\begin{document}


\maketitle

\begin{abstract}
The aging phenomena occurring in the course of cycling processes in the insertion host-guest compounds are discussed in the framework of a simple model. It takes into account two types of effects. One can be attributed to modifications of the host/guest solution interface in a form of an effective energetic barrier. The other
is associated with the host matrix disordering that is  
 translated into a change in the distribution of the host site energies as a function of the applied potential or the concentration of the guest species.   
It is found that the aging properties depend on the preparation mode,
the cycling conditions  and the insertion induced transformations. 
The impact of these transformations on the aging is determined by the matrix sensibility. 
\end{abstract}

\vspace{1cm}

\vspace{0.5cm}

\section{Introduction}
Insertion processes have received considerable interest because 
 of their applications in a  variety of technologically important 
 domains, e.g., storage devices \cite{Zuttel,Matsuda}
 high capacity batteries, electro-chromic devices, solar 
 cells, etc (see ref.~\cite{review,review1,review2} for a review).  These electro-active materials 
 are usually produced by insertion or doping of a host matrix by 
 neutral or charged guest species. Typical examples are the 
 intercalation compounds or conducting polymers. In many cases (e. 
 g. amorphous/porous materials or polymer films) the host matrix is 
 spatially or energetically disordered. This has some advantages \cite{Julien} (e.g., higher 
 capacity) in comparison to crystalline hosts. In particular, this allows 
 one to avoid the macroscopic insertion-induced phase transformations, such 
 as restructuring (crystalline $WO_3$) or staging (graphite). In its turn, 
 this broadens the voltage and composition ranges where a stable cycling can 
 be maintained without accumulating significant stress and strain field which affect
 the mechanical stability.

From a theoretical point of view the host matrix is usually described as a network which is characterized by a distribution of some relevant quantity, such as pore sizes or site
energies \cite{Gelb,Kudo,Stromme}. Then the observable insertion isotherm is represented
as an average over the host fluctuations. 
Despite a considerable progress \cite{Gelb} in the field,  
the consequences of the  disordered host morphology are 
still poorly understood. 
The problem is complicated by the structural changes occurring upon 
insertion \cite{Pablo,Ord1} or intercalation \cite{Prejean,BisqL}. In many 
cases mechanical strain generated in the intercalation/ deintercalation 
cycles induces structural as well as volume changes in the host 
material, leading to fracturing, cracking and even crumbling, and thereby to 
the host matrix irreversibility upon the cycling. One of the most challenging issues 
in the development of devices based on lithium intercalation materials is 
the control over structural changes and deformations produced by lithium 
insertion/removal. These can be manifested as spatially distributed internal 
distortions which lead either to topotactic insertion \cite{Julien} or even 
to macroscopic (up to $10\%$) volume dilatations \cite{BisqL,diffusion}. 
The latter are easily detectable (e.g., by optical 
profilometry \cite{BisqL}), while the internal distortions are rather 
difficult to detect and control. For this reason quite often the internal host 
structure remains poorly characterized. On the other hand, a 
coupling between the insertion and the dilatation modes has been 
shown \cite{JCPcomp,JPCcomp} to be responsible for well-pronounced thermodynamic 
features.  Even in the absence of detectable volume variations, the internal 
host distortion should contribute \cite{acta-info} a fluctuational term to 
the insertion thermodynamics. Therefore, one should be able (at least in 
principle) to distinguish these two effects. 
In the absence of a detailed microscopic information one has to analyze
the underlying physics directly from the experimental data and a plausible
insertion model. The insertion induced structural transformations of the host matrix can be translated into a probability distribution of the host-guest interaction energy that is conditional to the external potential or to the insertion level \cite{eactaflex}. Such an approach has recently been developed \cite{condpol} in application to the polaron energy distribution in conducting polymers . 

It is well known that  the life time of contemporary insertion systems mentioned above is limited by the aging processes. In fact, this is a superposition of various complex processes running at different time scales. From a quite general point of view the aging is equivalent to a degradation of the whole system involving its active and inactive components (see \cite{Vetter} for a recent review). Physically the aging is associated with an irreversible loss of capacity in accommodating the guest species in the course of cycling (charge-discharge,  adsorption-desorption)  or self-discharge processes. In the context of electrochemical intercalation studies several mechanisms are being discussed in the current literature.
Namely, the role of a degradation of active materials, changes in the electrode compounds, decomposition of electrolytes and formation of electrode - electrolyte interfaces are considered \cite{Vetter}.
It is clear that all these processes fall into two groups. One of them includes the interfacial properties. Namely, the formation and evolution of the electrode-electrolyte interface. This aspect has recently been analyzed theoretically \cite{Safari} within a multimodal aging model.  
The other point is related to the disordering phenomena in the bulk of the active material.
Theoretical modelling of this aging source is still lacking.
In this respect, based on our previous studies \cite{eactaflex,condpol},  we propose a theoretical approach that takes into account these two factors in a simple manner. More specifically, a superposition of the host matrix disordering and the loss of its activity during the cycling is considered.  
Recent experimental data on the capacity loss in both commercial and laboratory electrochemical cells are analyzed in this context.

\section{Background and the insertion model}
Let us outline first some basic principles. A typical cell consists of two electrodes (cathode and anode) with an electrolyte solution in between. The positive electrode (the cathode) is usually a lithiated metal oxide (for instance $LiCoO_2$), the negative electrode (the anode) is typically a layered graphite ($C$) matrix. Upon application of an external potential $V$ the following reactions take place
$$
LiCoO_2\to Li_{1-x}CoO_2+xLi^{+}+xe^{-},
$$
$$
C+xLi^{+}+xe^{-}\to CLi_x,
$$
where $x$ is the concentration of $Li$ in one of the electrodes.     
Therefore, during the charge or discharge processes the ions ($Li^+$, for instance) are transferred from one electrode to the other, while their charge is compensated by electrons passing through the external circuit under the applied potential $V$. In this way after the charge or discharge the neutral guest species appear inserted (intercalated as in the case of layered materials) in one of the host matrices. Then the work done by the electrons $eV$ is related to the change in the guest chemical potential (free energy per particle)
\begin{equation}
\mu_c-\mu_a=-eV,
\end{equation}
where $\mu_c$ and $\mu_a$ are the chemical potentials of neutralized ions in the cathode and anode, respectively. It is supposed \cite{review} that one of the counterparts ($\mu_a$ for instance) 
can be assumed constant $\varepsilon$, depending on whether the charge or discharge is considered.
As we will see later, the value of $\varepsilon$ might depend on the number of cycles reflecting the evolution of the electrode-electrolyte interface that acts as an energetic
barrier for the guest species.  
Thus one deals with the chemical potential $\mu$ of the guest that is related to the cell voltage
as $\mu=eV-\varepsilon$. Therefore, measuring the cell potential gives the information on the the chemical potential $\mu$ of the guest species. As the inserted ions are neutralized by electrons, $\mu$ can be decomposed into two parts \cite{review1}
\begin{equation}
\mu=\mu_i+\mu_e,
\end{equation}
where $\mu_i$ and $\mu_e$ are the chemical potentials of ions and electrons, respectively.
Although such a decomposition is well accepted, it is not unique because the ion-electron interaction inside the host could be partitioned in some arbitrary way between the two terms.
As it is argued \cite{review1}, in metallic compounds it is often possible to arrange the interaction terms so that $\mu_e$ is constant (e.g. the Fermi level). 
Then we are left with the chemical potential $\mu$  that is proportional to the applied potential $V$ and, on the other hand, it is a function of the guest species concentration $x$.

If $Q(t)$ is the charge transferred by electrons (that  is equal to the guest charge inside the matrix) as a function of time $t$, then the charge or discharge
current $I$ measured experimentally is given by 
\begin{equation}
I=\frac{\partial Q}{\partial t}=\frac{\partial Q}{\partial \mu}\frac{\partial\mu}{\partial t}=
Q_0 \frac{\partial x}{\partial \mu}\frac{\partial V}{\partial t}=
Q_0 C(\mu)\frac{\partial V}{\partial t}
\end{equation}
In what follows partial derivatives of the type $\partial F/\partial y$ mean that all other parameters on which the quantity $F$ might depend are fixed (e.g. temperature). 
Here $Q=Q(x)=Q_0x$ with $Q_0$ being the charge at $x=1$. At a constant voltage rate
$\partial V/\partial t=const$ (the so-called linear sweep voltammogram) the current
is proportional to the capacitance $C(\mu)=\partial x/\partial \mu$.
This is quite important relation between the observable quantity $I$ and the thermodynamics
$C(\mu)$ of the guest species inside the cell. Making some plausible or model assumptions on the $x$ vs $\mu$ behavior inside the cell (or at least in one of the electrodes) we can predict the output $I-V$ characteristics which determines the cell performance.       

In the context of theoretical studies the host matrix is usually represented as a network of sites where the guest species can be accommodated. Although the phenomena occurring near
the cathodic and anodic part could involve different dominating processes, in any case   
the ionic charge inside the matrix is compensated
by the electrons. For that reason the electrochemical intercalation is  similar
(at least, in some aspects) to a three-dimensional adsorption of neutral species.
Therefore one can capture the physics of the cathode or anode insertion within a unified theoretical scheme involving parameters which are specific to a given situation. 
Based on this analogy the intercalation is traditionally
described within the lattice gas (LG) model. In this approach the concentration $x$ of the guest species is related to the chemical potential $\mu$. In the mean-field approximation
we arrive at the well-known\cite{review1} Frumkin isotherm
\begin{equation}
\mu=\varepsilon+Wx+kT\ln{\left(\frac{x}{1-x}\right)}.
\label{frumkin}
\end{equation}
Here $\varepsilon$ is the host-quest interaction energy that depends on the nature of the host and guest species and on the properties of the host-solution interface. The second term corresponds to the guest-guest interaction inside the host matrix. The third term in (\ref{frumkin}) gives the entropy contribution to the overall energetic cost. In reality
the external potential $V$ (and consequently $\mu$) varies in a given finite range. Therefore,  the complete intercalation or deintercalation  is practically impossible because $x=0$ or
$x=1$ would require the limits $\mu=-\infty$ or $\mu=+\infty$, respectively. Keeping this fact in mind we could, however, accept such limits in order to obtain analytical estimations which do not alter our results for finite $\mu$. 

Since the controlling parameter is the chemical potential $\mu$ it is convenient to invert (\ref{frumkin}) in order to express the conditional
guest concentration $x=x(\mu|\varepsilon)$ as a function of $\mu$.
In the absence of interaction between the guest particles ($W=0$) the insertion isotherm obtained from (\ref{frumkin}) is given by the Langmuirian form
\begin{equation}
\label{xl}
x(\mu|\varepsilon)=\frac{\exp{[\beta(\mu-\varepsilon)}]}{1+\exp{[\beta(\mu-\varepsilon)}]},
\end{equation} 
where $\beta=1/kT$ is the inverse temperature. 

Another important point is that the host structure is never perfect. Even in the case of crystalline matrices there is always some fraction of defects (vacancies, dislocations, etc.)
which accumulates during the cycling.  
This means that in general the host matrix sites are not equivalent because of their local environment. For disordered (e.g. amorphous) matrices this feature is essential as it reflects the way of their generation.      

Because of the structural disorder in the host matrix the interaction energy is a fluctuating quantity distributed according to a probability density $f(\varepsilon)$. 
Then the observable isotherm $x(\mu)$ is an average over the host fluctuations
\begin{equation}
\label{x}
x(\mu)=\int d\varepsilon f(\varepsilon) x(\mu|\varepsilon) 
\end{equation}
This determines the differential capacitance $C(\mu)$
\begin{equation}
\label{C}
C(\mu)=\frac{\partial x(\mu)}{\partial \mu}=
\int d\varepsilon f(\varepsilon) \frac{\partial x(\mu|\varepsilon)}{\partial \mu}
\end{equation}
In what follows we are working with the dimensionless quantity $\mu=\beta \mu$ which makes the capacitance also dimensionless.

\section{Insertion induced transformations in a single cycle}
As is discussed above, the aging is related to the insertion induced transformations which are accumulated in the course of cycling processes. Let us first outline what happens in a single cycle. The host reaction to the insertion is accompanied by morphological changes of the matrix (e.g. doping induced modifications in conductive polymers). In our language this means that the distribution is conditional to the applied potential $\mu$ or to the insertion level $x(\mu)$. In what follows we discuss the effects which can arise from such a dependence.
In the case when the host energy distribution depends on the external potential $f(\varepsilon)=f(\varepsilon|\mu)$ we have two quasi-independent processes. One is the variation of the host morphology (translated into $f(\varepsilon|\mu)$), the other is the change in the insertion level $x(\mu)$ given by
\begin{equation}
\label{xm}
x(\mu)=\int d\varepsilon f(\varepsilon|\mu) x(\mu|\varepsilon)
\end{equation}
The additivity of these two processes is clear from the differential capacitance
\begin{equation}
\label{Cm}
C(\mu)=\frac{\partial x(\mu)}{\partial \mu}=
\int d\varepsilon f(\varepsilon|\mu) \frac{\partial x(\mu|\varepsilon)}{\partial \mu}
+
\int d\varepsilon \frac{\partial f(\varepsilon|\mu)}{\partial \mu} x(\mu|\varepsilon)
\end{equation} 
Here the first term on the r.h.s. is an analog of (\ref{C}). It corresponds to a tendency to increase the guest concentration with increasing external potential $\mu$. In fact, it is a conditional capacitance averaged over the host energy fluctuations. The second term corresponds to the variation of the host statistics with the applied potential.  This variation is a trace of the insertion induced matrix modifications (such as varying morphology in the form of internal distortions, local conformations, etc). 
It is to be noted that $C(\mu)$ given by (\ref{Cm}) is a combined quantity. 
The first term is strictly positive because of the thermodynamic stability requirements. In contrast, the sign and the magnitude of  the second term is not restricted by any fundamental law. It depends on the way the matrix is generated and on its ability to react on the variation of the potential $\mu$. Possible consequences of such a host reaction to the applied potential is analyzed in our previous studies \cite{JCPcomp,JPCcomp}. In what follows we are focusing on a coupling of the insertion induced effects for a given $\mu$ and the number of cycles $N$.

\section{The cycling}
As is mentioned above the overall cell performance is limited by both the anodic and the cathodic parts. Although the dominating processes occurring near each of them (interface evolution or restructuring) are different, in any case we deal with the insertion into
a host matrix. Therefore we are trying to develop a unified scheme applicable to both counterparts with changing the appropriate parameters.    
The matrix energy distribution is modeled as the following gaussian form
\begin{equation}
f_N(\varepsilon|\mu)=\frac{1}{\sqrt{2\pi s^2(\mu,N)}} 
\exp\left[-\frac{(\varepsilon-{\overline{\varepsilon}}(\mu,N))^2}{2s^2(\mu,N)}
\right]
\label{gmu}
\end{equation}
In this way we can take into account two types of effects. One of them is the modification of the
mean host-guest interaction ${\overline{\varepsilon}}(\mu,N)$ with the number of cycles $N$ and the external potential $\mu$.
This reflects the formation and evolution of the host-solution (or electrode-electrolyte) interface that acts as an energetic barrier preventing the penetration of the guest particles into the host matrix. Following the phenomenology discussed in the current literature \cite{Vetter,Ramos-shift}
we may accept that
the barrier becomes higher with increasing $N$.  In the simplest way this can be modeled as
\begin{equation}
{\overline{\varepsilon}}(\mu,N)=\varepsilon_0+\alpha(\mu) N =
\varepsilon_0 \left[
1+\frac{\alpha(\mu)}{\varepsilon_0} N
\right],
\label{ebar}
\end{equation}
where $\varepsilon_0$ is the initial value and the dimensionless parameter $\alpha^*=\alpha (\mu)/\varepsilon_0$ gives an increase of the barrier in a single cycle. In the context of lithium-ion batteries this effect is considered as dominating for the aging of the anodic part \cite{Vetter,Aurbach}. It is also argued \cite{Vetter} that in practice it is almost always possible to find an
anode-solvent couple that would be reasonably stable during the life time of the cell. Therefore, in what follows we assume that $\alpha(\mu)=\alpha$ is a constant independent of the
external potential.  

Then the evolution of the cathodic part is the time limiting process than is responsible for the performance of the whole system.  
In this case the role of the interface is thought to be negligible ($\alpha \to 0$) while the dominating 
effect is related to the insertion induced host restructuring. This can be taken into account as
a change of the distribution width
\begin{equation}
s(\mu,N)=\sigma_0+\gamma N \sigma(\mu)=\sigma_0 \left[1+ \gamma  \frac{\sigma(\mu)}{\sigma_0}N\right],
\label{s}
\end{equation}
where $\sigma_0$ is the initial width that depends on the host sample preparation.
The second term takes into account the host-guest coupling in the course of cycling while
the $\mu$ dependent term $\sigma(\mu)$ corresponds to the insertion induced matrix modifications in a single cycle. Here $\gamma$ is a cycling parameter that, for a given material, depends on the external conditions (e.g. the cycling rate, the charge/discharge depth, temperature, etc.). For a given $\mu$ the dimensionless parameter $\gamma^*=\gamma \sigma(\mu)/\sigma_0$ gives the capacity loss in a single cycle.    

In what follows we consider the well-accepted case
when the distribution becomes broader (disordering) in the course of insertion. A typical example is the doping induced broadening of the polaron energy distribution in conducting polymers films  \cite{condpol}. There could be various mechanisms of the broadening (the film morphology or/and polymer conformation variations, formation of traps and disordered dipolar type arrays, etc.). For the moment, without resorting to a concrete model one can easily imagine that the width varies smoothly from some initial value $\sigma_0$  at low insertion level ($\mu \to -\infty$) to some final value $\sigma_0+\gamma N$ corresponding to a saturation at ($\mu \to \infty$ ). This can be modeled by introducing the following modulation function
\begin{equation}
\sigma(\mu)=\frac{1}{2}
\left[
1+erf{\left[
\frac{\sqrt{2}}{2\delta} (\mu-\mu_0)
\right]}
\right].
\end{equation}
Therefore, the cycling conditions $\gamma^* N$  control the magnitude of the width variation, $\delta$ determines the smoothness and the point of inflection is given by $\mu_0$. 
As is discussed above, the host matrix restructuring is translated into the energetic disorder.
Thus $\mu_0$ can be interpreted as an energetic scale at which the matrix response becomes well detectable, with $\delta$ giving the response intensity.  
Since we focus on the the insertion induced effects, it is reasonable to accept the zero-temperature ($\beta \to \infty$) approximation for the conditional isotherm (\ref{xl})
\begin{equation}
\label{zt}
x(\mu|\varepsilon)= H(\mu-\varepsilon), 
\end{equation}  
where $H(\cdots)$ is the Heaviside step function. Then the capacitance (\ref{Cm}) becomes
\begin{equation}
C(\mu,N)=f_N(\mu|\mu)+ \int_{-\infty}^{\mu} d \varepsilon \frac{\partial f_N(\varepsilon|\mu)}{\partial \mu}
=
f_N(\mu|\mu)\left[
1-\gamma N\frac{\mu-\varepsilon_0-\alpha N}{\sigma_0+\gamma N\sigma(\mu)}\frac{\partial \sigma(\mu)}{\partial \mu}
\right]
\label{Cms}
\end{equation} 
In fact we have a superposition of two effects. First is the broadening of the capacitance curve that is gouverned mainly by the $f_N(\mu|\mu)$ term that is a gaussian centered around $\mu=\varepsilon_0$. The second is the modulation induced by the $\sigma(\mu)$ dependence that is also reduced to a gaussian form
centered around $\mu=\mu_0$. Thus, the shape of the capacitance curve depends on the
distance between $\mu_0$ and $\varepsilon_0$ and the position of the applied potential $\mu$
with respect to these two characteristic energy scales.  Therefore, as it analyzed in \cite{eactaflex,condpol}, the insertion induced matrix modifications are associated with well-detectable features in the capacitance curve. In what follows we are focusing on 
the behavior of the "principal" peak at $\mu=\varepsilon_0$ with the number of cycles. 

The evolution of the  capacitance curves with the number of cycles $N$ is illustrated in Fig.~1.
Two effects are clearly seen. One of them is a shift of the curves towards higher external potentials $\mu$. Such a shift is well documented in the current literature (see for example
recent results \cite{Ramos-shift} on thin films of $LiCoO_2$). 
This process (governed by the parameter $\alpha)$ is a consequence of the energetic barrier mentioned above. Physically this means that the insertion becomes more energetically costly with the cycling. In reality one is working in a given range of the external potentials (e.g. pressure or voltage "windows"). Therefore, the shift is equivalent to a gradual decrease of the capacitance within the window. This can be seen through the evolution
at the point $\mu=\varepsilon_0$ where $C(\mu,N)$ calculated for the "fresh" system ($N=0$) has a gaussian peak. The second effect is a broadening of the curves with increasing $N$. This process is associated with a disordering of the host matrix during the cycling accompanied by the capacity loss. In order to estimate the loss let us focus again on the initial peak at $\mu=\varepsilon_0$ and ignore the shift ($\alpha=0$). 
Then from equation (\ref{Cms}) we obtain
\begin{equation}
C(\mu=\varepsilon_0,N)=
\frac{1}{\sqrt{2\pi[\sigma_0+\gamma N \sigma(\mu=\varepsilon_0)]^2}}=
\frac{C_0}{1+\gamma N\sigma(\mu=\varepsilon_0)/\sigma_0},
\label{Cmss}     
\end{equation}
where $C_0=1/\sqrt{2\pi\sigma_0^2}$ is the initial ($N=0$) capacitance.
As is noted above, the dimensionless parameter $\gamma^*=\gamma \sigma(\mu=\varepsilon_0)/\sigma_0$ is related to the capacity loss in a single cycle.
Thus we arrive at the following simple form
$$
C(N)=\frac{C_0}{1+\gamma^{*}N}
$$
that has been used for the fitting to experimental data \cite{Mehrens} on cycling in doped lithium-nickel-cobalt oxides. The results are presented in Fig.~2. It is seen that a $1/(1+\gamma^{*}N)$ dependence fits quite well to the data. Obviously there is a lot of phenomena hidden behind the fitting parameter $\gamma^*$ (in our case one of them is the doping level). Nevertheless it is interesting to note that our estimation of the capacity loss per cycle ($\gamma^* \approx 0.1 \%$) is in agreement with that found in \cite{JulienI} for disordered $MoS_2$.   

Therefore, the aging properties depend on several factors: the preparation mode ($\sigma_0$, $\varepsilon_0$),
the external conditions ($\gamma$) and the insertion induced transformations ($\sigma(\mu)$). 
The impact of these transformations on the aging is determined by the matrix sensibility ($\delta$) but also by an interplay of two energetic scales ($\mu_0$ and $\varepsilon_0$).
Conventionally the end of life is defined as when after $N_c$ cycles the system reaches certain "critical" percentage $K$ of its initial capacity, that is, $C(N_c)/C_0=K$. The magnitude of $K$ is system dependent. For instance, in the case of lithium-ion batteries it is accepted \cite{Spotnitz} that $K=0.8$. From this criterium we can estimate the relative distribution width as
\begin{equation}
\frac{s(\mu=\varepsilon_0,N_c)-\sigma_0}{\sigma_0}=
\gamma N_c\frac{\sigma(\mu=\varepsilon_0)}{\sigma_0}=\frac{1}{4}
\end{equation}  
Therefore, the aging does not require significant disorder variations in the host matrix (about 25\%). This points out towards a necessity of designing structurally stable host matrices in order to avoid significant cycling induced transformations.   

It is also instructive to analyze the case of low initial disorder ($\sigma_0 \to 0$) which
corresponds to a nearly crystalline host matrix. In this case from eq. (\ref{Cmss}) we arrive at the following limit
\begin{equation}
C(\mu=\varepsilon_0,N)=
\frac{1}{\sqrt{2\pi}}\frac{1}{\gamma N\sigma(\mu=\varepsilon_0)}.
\label{Cmsl}     
\end{equation} 
This result suggests that the aging process is practically independent of the matrix preparatioin ($\sigma_0$), as it should be expected for ordered (e.g. crystalline) materials.
Then the capacity loss in the course of the cycling is determined by the host reaction
$\sigma(\mu)$ and the cycling conditions ($\gamma$).
 
In order to verify that the features mentioned above are not specific to the step-wise approximation (\ref{zt}) accepted for $x(\mu|\varepsilon)$ we have taken the conditional isotherm in the form (\ref{xl}), while the host distribution is a uniform step-wise
\begin{equation}
f_N(\varepsilon|\mu)=\frac{1}{2s(\mu,N)} H[\varepsilon+s(\mu,N)]\cdot H[s(\mu,N)-\varepsilon]
\end{equation}
such that the host-guest interaction energy $\varepsilon$ is uniformly distributed between
$\varepsilon=-s(\mu,N)$ and $\varepsilon=s(\mu,N)$ with the mean value $\varepsilon_0=0$
and the width given by $2s(\mu,N)$.
In this case the isotherm (\ref{xm}) becomes
\begin{equation}
x(\mu,N)=\frac{1}{2s(\mu,N)} 
\ln {\left[
\frac{1+\exp{(\mu+s(\mu,N))}}{1+\exp{(\mu-s(\mu,N))}}
\right]} 
\end{equation}  
The corresponding capacitance curve (\ref{Cm}) exhibits qualitatively similar features (specific maxima, plateaus and minima) which are directly related to the variation of the distribution
width $s(\mu,N)$ with the applied potential $\mu$ and the number of cycles $N$.

\section{Conclusion}
The aging phenomena occurring in the course of cycling processes in the insertion host-guest compounds are discussed in the framework of a simple model. It takes into account two types of effects. One of them can be attributed to modifications of the host/guest solution interface in a form of an effective energetic barrier. The other
is associated with the host matrix disordering that is  
 translated into a change in the distribution of the host site energies as a function of the applied potential or the concentration of the guest species. It it is demonstrated that in a single cycle the insertion-induced transformations of the host matrices are well detectable in the form of characteristic peaks, minima and plateaus appearing in the capacitance curves. 
In particular, it is shown that the shape of the capacitance curve is quite sensitive to the characteristic energy scales $\varepsilon_0$ and $\mu_0$ giving the energy costs for the insertion and the matrix transformation, respectively. 

It is found that the aging properties depend on several factors: the host preparation mode,
the cycling conditions  and the insertion induced transformations. 
The impact of these transformations on the aging is determined by the matrix sensibility but also by an interplay of two energetic scales ($\mu_0$ and $\varepsilon_0$).
Recent experimental data on the capacity loss in electrochemical cells are analyzed in this context. It is shown  that our estimation of the capacity loss per cycle ($\gamma^* \approx 0.1 \%$) is in agreement with that found in \cite{JulienI} for disordered $MoS_2$.   

It should be noted that the present analysis is applicable to systems operating at nearly equilibrium conditions. Therefore, all possible kinetic limitations (e.g. the charge/discharge rate and depth) are not considered explicitly. In the simplest way these effects can be incorporated in our formalism as the Butler-Volmer reaction with fluctuating rate constants.
This would allow us to analyze more realistic situations (close to industrially important cases). This issue is left for future studies.   


\newpage

\begin{figure}
\caption{Relative capacitance $C(N)/C_0$ as a function of the applied potential $\mu$.\newline
Part (a) ("shift"): $\sigma_0=1, \varepsilon_0=1, \delta=0.3, \mu_0=1$, $\alpha=0.01$, $\gamma=0$,$N=0$ (solid), $N=50$ (dash)
\newline
Part (b): ("broadening")
$\sigma_0=1, \varepsilon_0=1, \delta=0.3, \mu_0=1$, $\alpha=0$, $\gamma=0.01$,$N=0$ (solid), $N=20$ (dash)
All quantities are dimensionless.}
\end{figure}

\begin{figure}
\caption{Relative capacitance as a function of the cycle number at different doping levels.  
All quantities are dimensionless.}
\end{figure}
\end{document}